\documentclass[runningheads]{llncs}
\usepackage{graphicx}
\usepackage{array}
\usepackage{amssymb}
\usepackage{bm}
\usepackage{booktabs}
\usepackage{cite}
\usepackage{footnote}
\usepackage{graphicx}
\usepackage[T1]{fontenc}
\usepackage{multirow}
\usepackage{subfigure}
\usepackage{url}
\usepackage[misc]{ifsym}

\begin{document}
\title{AutoSUM: Automating Feature Extraction and Multi-user Preference Simulation for \\
Entity Summarization}
\titlerunning{AutoSUM}
%
\author{Dongjun Wei\inst{1,2} $^{\star}$  \and
Yaxin Liu\inst{1,2} \thanks{Equal contribution. Fuqing Zhu is the corresponding author.} \and
Fuqing Zhu\inst{1}   
\and
Liangjun Zang\inst{1} \and \\
Wei Zhou\inst{1} \and
Yijun Lu\inst{3} \and
Songlin Hu\inst{1,2}}
\authorrunning{Wei D, Liu Y, et al.}
%
\institute{
Institute of Information Engineering, Chinese Academy of Sciences, China \and
School of Cyber Security, University of Chinese Academy of Sciences, China \and
Alibaba Cloud Computing Co. Ltd., Beijing, China\\
\email{\{weidongjun, liuyaxin, zhufuqing, zangliangjun, zhouwei, husonglin\}@iie.ac.cn} \\
\email{yijun.lyj@alibaba-inc.com}}
\maketitle              
\begin{abstract}
With the growth of knowledge graphs, entity descriptions are becoming extremely 
lengthy. Entity summarization task, aiming to generate diverse, comprehensive and representative 
summaries for entities, has received an increasing interest recently. In most previous methods, 
features are usually extracted by the hand-crafted templates. Then the feature selection and 
multi-user preference simulation take place, depending too much on human expertise. 
In this paper, a novel integration method called AutoSUM is proposed for automatic feature extraction and 
multi-user preference simulation to overcome the drawbacks of previous methods.
There are two modules in AutoSUM: 
extractor and simulator. The extractor module operates automatic feature extraction based on a BiLSTM 
with a combined input representation including word embeddings and graph embeddings. Meanwhile, 
the simulator module automates multi-user preference simulation based on a well-designed 
two-phase attention mechanism (i.e., entity-phase attention and user-phase attention). 
Experimental results demonstrate that AutoSUM produces the state-of-the-art performance 
on two widely used datasets (i.e., DBpedia and LinkedMDB) in both F-measure and MAP.
The source code and outputs are available at \url{https://github.com/WeiDongjunGabriel/AutoSUM}.
\end{abstract}
\section{Introduction}\label{intro}
Semantic data enables users or machines to comprehend and manipulate the 
conveyed information quickly~\cite{Gunaratna2016GleaningTF}. In major
knowledge graphs, semantic data describes entities by 
Resource Description Framework (RDF) triples, referred as triples~\cite{Cheng2011RELINRA}.
With the growth of knowledge graphs, entity descriptions are becoming extremely 
lengthy~\cite{Thalhammer2016LinkSUMUL}.
Since Google first released the knowledge graph,
``get the best summary'' for entities has been one of the main contributions 
in Google Search\footnote{\url{https://www.google.com}}~\cite{Thoma:2019:FED:3313948.3306128}. 
Specifically, 
Google Search returns a top-$k$ subset of triples 
which can best describe the entity from a query on the right-hand side of the result pages~\cite{Liu2019EntitySS}.
Motivated by the success of Google Search, entity summarization task has received an increasing 
interest recently~\cite{GOBlog, Thoma:2019:FED:3313948.3306128}, it aims to generate diverse, 
comprehensive and representative 
summaries for entities. In addition, entity summarization has been integrated into various applications
such as document browsing, Question Answering (QA), \emph{etc}~\cite{Liu2019EntitySS}.

Most previous entity summarization methods are adopted from
random surfer~\cite{Cheng2011RELINRA}, clustering~\cite{Gunaratna2015FACESDE,Gunaratna2016GleaningTF} 
and Latent Dirichlet Allocation (LDA)~\cite{Pouriyeh2017ESLDAES}
models, depending too much on the hand-crafted templates 
for feature extraction as well as human expertise for feature selection.
Meanwhile, entities are capable to represent diverse information (or multi-aspect information)
in knowledge graphs~\cite{Sydow2010DIVERSUMTD}, resulting in different user preference (sometimes 
multi-user preference~\cite{wang2019user}). 
Take entity \emph{Triathlon\_ at\_the\_2000\_Summer\_Olympics\_Men's}
in DBpedia\footnote{\url{https://wiki.dbpedia.org}} for instance,
different users may prefer to the \emph{medal}, \emph{event} or \emph{type} of
this entity, respectively.
In order to generate more diverse summaries, 
the specific model needs to be selected for providing a
more distinguishable multi-user preference simulation~\cite{Gunaratna2015FACESDE, Sydow2010DIVERSUMTD}.
However, due to the countless quantities and unpredictable types of entities in real large-scale 
knowledge graphs, extracting discriminative features or selecting suitable models 
based on human expertise could be arduous~\cite{Liu2019EntitySS}. 

In this paper, a novel integration method called AutoSUM is proposed for automatic feature extraction and 
multi-user preference simulation to overcome the drawbacks of above previous models.
There are two modules in AutoSUM: extractor and simulator. The extractor module operates automatic 
feature extraction based on a BiLSTM with a combined input representation including word embeddings 
and graph embeddings. Meanwhile, 
the simulator module automates multi-user preference simulation based on a well-designed 
two-phase attention mechanism (i.e., entity-phase attention and user-phase attention). 
Experimental results demonstrate that AutoSUM produces the state-of-the-art performance 
on two widely used datasets (i.e., DBpedia and 
LinkedMDB~\footnote{\url{http://data.linkedmdb.org}}) in both F-measure and MAP.

\section{Related Work}\label{work}
Previous entity summarization methods mainly rely on human expertise.
To find the most central triples,
RELIN~\cite{Cheng2011RELINRA} and SUMMARUM~\cite{Thalhammer2014BrowsingDE} 
compute the relatedness and informativeness based on the features
extracted from hand-crafted templates.
Meanwhile,  
FACES~\cite{Gunaratna2015FACESDE} and ES-LDA~\cite{Pouriyeh2017ESLDAES} introduce 
a clustering algorithm and LDA model for 
capturing multi-aspect information, respectively.
In order to generate more diverse summaries, 
the specific models need to be selected for providing a
more distinguishable multi-user preference simulation~\cite{Gunaratna2015FACESDE, Pouriyeh2017ESLDAES}.
However, due to the countless quantities and unpredictable types of entities
in the real large-scale knowledge graphs, extracting discriminative features 
and selecting suitable models
based on human expertise could be arduous.

Recently, 
deep learning methods relieve the dependency on human expertise in
Natural Language Processing (NLP)~\cite{Luong2015EffectiveAT} community.
To generate the summaries without human expertise,
an entity summarization method with a single-layer attention (ESA)~\cite{Wei2019ESAES} is proposed to
calculate the attention score for each triple.
Then top-$k$ triples which have the highest attention scores are selected 
as the final results. 
However, ESA cannot extract features and capture multi-aspect information with 
the single-layer attention mechanism. 
Following ESA work, our proposed AutoSUM automates feature extraction and 
multi-user preference 
based on a novel extractor-simulator structure. In extractor, 
a BiLSTM with a combined input representation is utilized for feature extraction. 
The word embeddings and graph embeddings are included.
Meanwhile, in simulator, a two-phase attention mechanism is designed 
for multi-user preference simulation.

\section{Proposed Model}\label{model}
\subsection{Problem Description}
An RDF triple is composed of a subject, a predicate, and an object. 
In major knowledge graphs, an entity of which is then defined as a subject
with all predicates and corresponding objects to those predicates.
When a user queries an entity in a knowledge graph, 
a set of triples $\left\{ t_{1},t_{2},\cdots,t_{n} \right\}$ related with the entity will be returned, 
referred as an entity description document $d$, where $t_i$ is the $i$-th triple in $d$.
Following Google Search~\cite{Liu2019EntitySS, GOBlog}, given a positive integer $k$,
the summary of an entity is a top-$k$ subset of $d$ which can best describe the entity. 
\subsection{Overview}
As shown in Figure 1, AutoSUM has a novel extractor-simulator structure.
The extractor extracts the features of triples in $d$
as $ h=\left\{ h_{1}, h_{2}, \cdots, h_{n} \right\} $, where $h_i$ is the feature vector of $t_i$.
Given $h$, the simulator calculates the attention scores $a = \{a_1, a_2, \cdots, a_n\}$,
where $a_i$ is the attention score of $t_i$.
Then top-$k$ triples with the highest attention scores 
will be selected as the summary of an entity.
\begin{figure}\label{model-fig}
\centering
\includegraphics[scale=0.35]{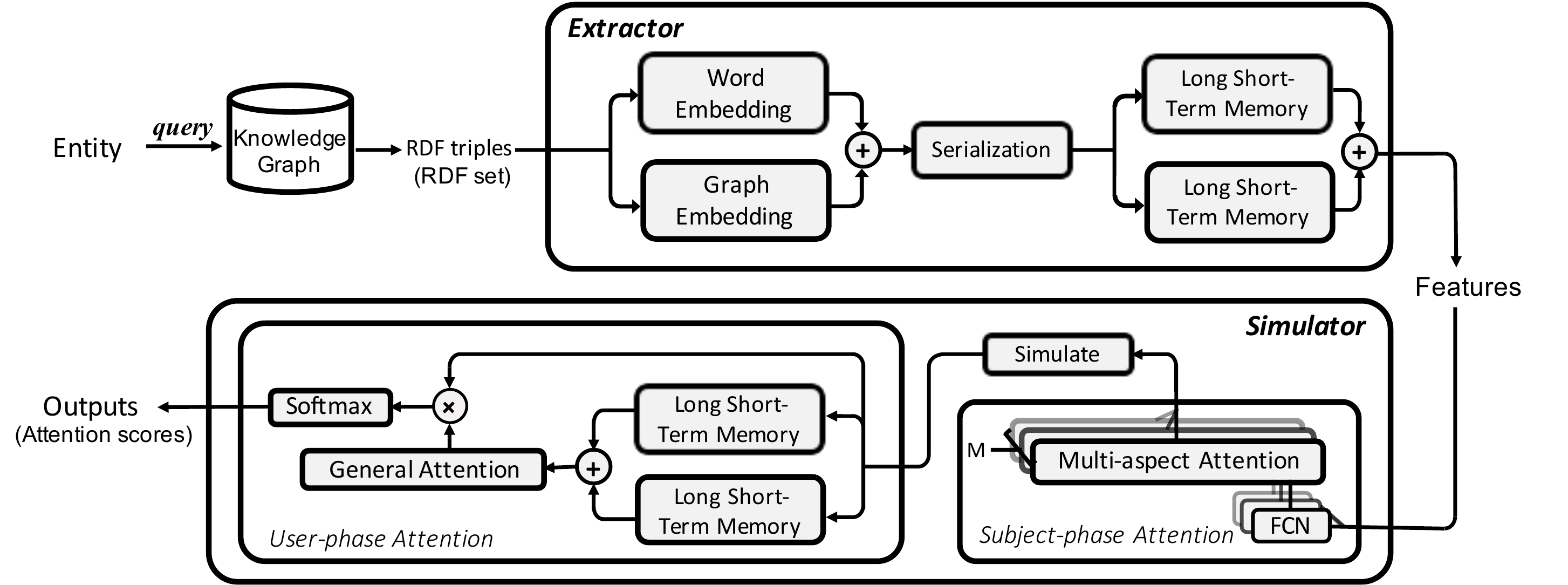}
\caption{The architecture of AutoSUM.}
\end{figure}

\subsection{Extractor}
The extractor module in AutoSUM aims at extracting features of triples automatically. 
In this section, we introduce the input representation and the automatic feature 
extraction in details.

\subsubsection{Input Representation}
As discussed above, the triples related with an entity share the same subject
with different predicates and corresponding objects to those predicates. 
In order to map predicates and objects into a continuous vector space for feature extraction,
we apply a combined input representation method including
word embeddings and graph embeddings.
Then we concatenate the embeddings of the predicates and corresponding objects
as the representation for each triple.

\textbf{Word Embedding:} 
Learning word embeddings has been an effective method to enhance the performance of entity sumamrizers. 
In ES-LDA$_{ext}$~\cite{Pouriyeh2018CombiningWE}, 
Pouriyeh \emph{et al.} stated the key point of learning word embeddings was the definition for ``words''. 
Following Pouriyeh's work, we extract predicates and objects of triples as our words. 
Take ``\emph{http://dbpedia.org/ontology/goldMedalist}'' for instance, 
we extract “\emph{goldMedalist}” as the word for the above predicate.
Given the embeddings of words, we then initialize a word embedding (lookup) table for future training.

\textbf{Graph Embedding:} 
Obviously, simple word embeddings cannot represent triples with a graph structure. 
To fully encode the graph information, we utilize a graph embedding technique called 
TransE~\cite{bordes2013translating} to pretrain the whole knowledge graph in the dataset. 
Given the embeddings of tirples, we then initialize a graph embedding table for future training.

\subsubsection{Automatic Feature Extraction}
In Named Entity Recognition (NER) task, the bidirectional LSTM (BiLSTM) has been widely used for
automatic feature extraction~\cite{lample2016neural}. For instance, in order to automatically extract features from a 
small and supervised training corpus, 
an LSTM-CRF model was proposed by Lample \emph{et al.}~\cite{lample2016neural},
utilizing a BiLSTM for feature extraction and conditional random fields~\cite{Lafferty2001ConditionalRF} for 
entity recognition. 
The BiLSTM extracted representative and contextual features of a word, 
aligning with other words in the same sentence~\cite{Graves2013GeneratingSW}.
As for summarizing entities, we also apply a BiLSTM to extract features 
of a triple,
aligning with other triples related with the same entity.
Specifically, due to the uncertain timing sequence of triples, we first map 
(serialize) the triples into a sequence comes randomly. Then we feed the 
input representation of triples in the sequence to the BiLSTM, 
and take the outputs as the extracted features for those triples.
      
\subsection{Simulator}
The simulator in AutoSUM aims at simulating multi-user preference based on a well-designed two-phase attention
mechanism (i.e., entity-phase attention and user-phase attention). Entity-phase attention captures multi-aspect 
information from an entity, user-phase attention then simulates multi-user preference based on the captured information.
In this section, we present the details of entity-phase attention and user-phase attention.

\subsubsection{Entity-Phase Attention}
The intuition of entity-phase attention is straightforward.
Since the single-layer attention mechanism in ESA~\cite{Wei2019ESAES} cannot capture 
multi-aspect information,
we then design a multi-aspect attention mechanism with multiple (stacked) attention layers 
to overcome the drawback of ESA.
One seminal work using stacked attention layers is neural machine translation (NMT)~\cite{Luong2015EffectiveAT},
where the stacked attention layers (Transformer)~\cite{Vaswani2017AttentionIA} are utilized to 
capture the multi-aspect information from a sentence.
To our knowledge, we are the first to utilize the stacked attention layers to 
capture the multi-aspect information from an entity.
Specifically, different attention layers capture information from an entity in
different aspects. In each attention layer, a general attention function~\cite{Luong2015EffectiveAT}  
is utilized to calculate the relevance between each triple and
the information captured from the attention layer, termed attention scores.
Here, instead of combining all attention layers to generate overall attention scores of Transformer~\cite{Vaswani2017AttentionIA}, 
we directly output the attention scores from each attention layer 
for multi-user preference simulation in user-phase attention.
Notice that the number of the attention layers is a hyper-parameter 
which can be tuned during training.
      
\subsubsection{User-Phase Attention}
When users browse triples,
they will allocate high preference values (more attention) 
to triples which are more related with the information they are interested in~\cite{Gunaratna2015FACESDE}.
Meanwhile, as described above, entity-phase attention consists of different attention layers
for capturing information in different aspects. In each attention layer, 
a general attention function  
is utilized to allocate higher attention scores to the triples
which are more relevant to the information captured from the attention layer.
To simulate the preference of users who are
interested in the information captured by the current attention layer, 
user-phase attention assigns the user preference 
values of each triple with the same attention scores from the attention layer. 
Then different distributions of attention scores in different attention layers 
simulate the different preference of different users (multi-user preference).
          
After simulating the multi-user preference, we have to allocate
different attention scores for different user preference rather than treating them equally. 
The main reason is that some user preference may represent the preference of most users 
for an entity, while others may represent the preference of few users for the same entity. 
Allocating proper attention scores for each user preference is critical to generate a more
comprehensive entity summarization result. Therefore, we combine a BiLSTM with a general 
attention score function for allocation.
In NER, a BiLSTM can maintain the independence and capture the intrinsic 
relationships among words~\cite{Graves2013GeneratingSW}. Similarly, a BiLSTM is adopted in user-phase attention to preserve 
independence as well as capture the intrinsic relationships between different user preference.
Then the outputs of the BiLSTM are taken as the inputs to a general attention score function,
in order to allocate attention scores for each user preference. 
At last, we integrate all the user preference based on the allocated attention scores.
In addition, due to the uncertain order in user preference like triples, we 
also randomly map the user preference into a sequence as our input of the BiLSTM.

\subsection{The Complete Pipeline}
In this section, we demonstrate the complete pipeline of AutoSUM.
As described in Section 3.1, 
the input of AutoSUM is
an entity description document $ d = \{ t_{1}, t_{2}, \cdots, t_{n} \} $.
Here, $t_i$ is the $i$-th triple in $d$, which
is composed of a same subject $s$, a predicate $p_i$ and an object $o_i$.
Given $d$, we first split $d$ into a predicate 
set $p= \{ p_{1}, p_{2}, \cdots, p_{n} \} $ and an object set $o= \{ o_{1}, o_{2}, \cdots, o_{n} \}$,
respectively.  Given $p$ and $o$, we combine word embeddings and graph embeddings 
to map $p_{i}$ and $o_{i}$ into a continuous vector space and concatenate them as $e_{i}$, 
recursively. Given $e=\{ e_{1}, e_{2}, \cdots, e_{n} \}$, we randomly map $e$ into 
a sequence $q = ( q_{1}, q_{2}, \cdots, q_{n} )$. 
Then we apply a BiLSTM to extract the features vector $h_{i}$ of $q_{i}$ as follows,
\begin{eqnarray}
& \overrightarrow{h_{i}} = LSTM_{L}(q_{i}, \overrightarrow{h_{i-1}}), i \in [ 1, n ] , & \nonumber \\
& \overleftarrow{h_{i}} = LSTM_{R}(q_{i}, \overleftarrow{h_{i-1}}), i \in [ 1, n ] , & \nonumber \\
& h_{i}=[\overrightarrow{h_{i}}, \overleftarrow{h_{i}}], c = [ \overrightarrow{c}, \overleftarrow{c} ],
\end{eqnarray}
where $\overrightarrow{c}$ and $\overleftarrow{c}$ are the final hidden states in forward and 
backward LSTM networks. Given $h=\{ h_{1},h_{2},\cdots,h_{n} \}$ and $c$, we utilize the multi-aspect 
attention mechanism to capture multi-aspect information. Specifically, for the $j$-th attention layer 
in multi-aspect attention mechanism, we calculate the attention score $s_{j}^i$ for triple $t_{i}$ 
with a general score attention function as follows,
\begin{eqnarray}
& s_{j}^i = score_{j}(h_{j}, c)=h_{i}^T W_{j} c, i \in [1,n], j \in [1,m], & \nonumber \\
& s_{j}=[s_{j}^1, s_{j}^2, \cdots, s_{j}^n],
\end{eqnarray}
where $W_{j}$ is a parameter matrix of the general attention score function in the $j$-th attention 
layer, and $m$ is the number of attention layers in the multi-aspect attention mechanism. 
Given $s=\{s_{1}, s_{2}, \cdots, s_{m}\}$, we then simulate the preference of
the $j$-th user $u_j$ who is interested in the information of triple $t_{i}$ 
captured by the $j$-th attention layer as follows,
\begin{eqnarray}
& u_{j}^i = s_{j}^i, i \in [1,n], j \in [1,m], & \nonumber \\
& u_{j}=[u_{j}^1, u_{j}^2, \cdots, u_{j}^n],
\end{eqnarray}
where $u_{j}^i$ is the preference value allocated to triple $t_i$ by $u_j$.
Given $u =$ \textsl{\{} $ u_{1}, u_{2}, \cdots, u_{m} $ \textsl{\}}, we randomly map $u$ into a sequence 
$q^* = (q^*_{1}, q^*_{2}, \cdots, q^*_{m})$ and utilize a BiLSTM to encode $u_{j}$ into $u^{*}_{j}$ as follows,
\begin{eqnarray}
& \overrightarrow{u^{*}_{j}}=LSTM_{L}(q^*_{j}, \overrightarrow{u^{*}_{j-1}}),j \in [1,m], & \nonumber \\
& \overleftarrow{u^{*}_{j}}=LSTM_{R}(q^*_{j}, \overleftarrow{u^{*}_{j-1}}),j \in [1,m], & \nonumber \\
& u^{*}_{i} = [\overrightarrow{u^{*}_{j}}, \overleftarrow{u^{*}_{j}}], c^{*}=[\overrightarrow{c^{*}}, \overleftarrow{c^{*}}],
\end{eqnarray}
where $\overrightarrow{c^{*}}$ and $\overleftarrow{c^{*}}$ are the final hidden states from 
forward and backward LSTM networks. Then we calculate the attention score for user 
preference as follows,
\begin{equation}
a^{*} = [u^{*}_{1}, u^{*}_{2}, \cdots, u^{*}_{m}]W^{*}c^{{*}^T},
\end{equation}
where $W^{*}$ is a parameter matrix of the general attention score function.
Having obtained $a^{*}$, we integrate different user preference to generate the final attention 
score for each triple $t_i$ in $d$ as follows,
\begin{equation}
a = Softmax([u_{1}, u_{2}, \cdots, u_{m}]a^{*^{T}}) = [a_1,a_2,\cdots,a_n ].
\end{equation}
Finally, we employ cross-entropy loss and define the loss function $L$ for AutoSUM,
\begin{equation}
L(a, \overline{a}) = CrossEntropy(a, \overline{a}).
\end{equation}
Here, $\overline{a}=\{ \overline{a}_{1}, \overline{a}_{2}, \cdots, \overline{a}_{n} \}$ 
is a gold(real) attention score vector
associated with above entity from ESBM dataset.
Specifically, we count the frequency of the $i$-th 
triple $t_i$ selected by users in ESBM dataset following ESA work, denoted as $ c_{i} $. 
Then the gold attention score $\overline{\alpha}_{i}$ of $t_i$ is formulated as 
follows,
\begin{equation}
\overline{\alpha}_{i}=\frac{c_{i}}{\sum_{i=1}^nc_{i}}.
\end{equation}

\section{Experiments}\label{exp}
\subsection{Experimental Setup}
\subsubsection{Dataset}
In this paper, we utilize ESBM dataset v1.1,
which consists of $6.8$k triples related with $125$ entities from DBpedia~\cite{Bizer2009DBpediaA} 
and $2.6$k triples related with $50$ entities from LinkedMDB~\cite{Consens2008ManagingLD}.
Given an entity,
ESBM asks $5$ different users to select top-$5$ and top-$10$ triples 
which can best describe the entity.
In addition, ESBM provides an evaluator
for the comparison of different entity summarization methods.
Both datasets and evaluator
can be accessed from the ESBM website\footnote{\url{http://ws.nju.edu.cn/summarization/esbm/}}.

\subsubsection{Baselines}
Our baselines consist of some existing state-of-the-art entity summarization methods, 
including RELIN~\cite{Cheng2011RELINRA}, DIVERSUM~\cite{Sydow2010DIVERSUMTD}, CD~\cite{CD}, 
FACES~\cite{Gunaratna2015FACESDE}, LinkSUM~\cite{Thalhammer2016LinkSUMUL}, 
MPSUM~\cite{Wei2018MPSUM} and ESA~\cite{Wei2019ESAES}. MPSUM~\footnote{\url{https://github.com/WeiDongjunGabriel/MPSUM}} 
is an open source implementation of ES-LDA. 
To provide ablation studies, we also modify the original AutoSUM 
into $5$ different versions, denoted as AutoSUM$^{1\sim5}$, which will be futher illustrated in Section 4.3.

\subsubsection{Evaluation Methodology}
Summarization tasks can be mainly divided into 
extractive and non-extractive tasks~\cite{liu2018graph,ahmed2019data},
which orient to unstructured and structured data, respectively.
Sydow \emph{et al.}~\cite{Sydow2013TheNO} stated that entity summarization
task could be treated as an extractive task of information retrieval (IR).
IR returns the most relevant documents for a query,
while entity summarization selects the top-$k$ triples related with an entity.
Following previous work, we utilize F-measure and mean average precision (MAP) metrics
for evaluation, which are two standard evaluation metrics in IR~\cite{Hripcsak2005TechnicalBA,Liu2019EntitySS}. 
F-measure is the harmonic mean of recall and precision, and MAP is the mean average 
of precision. 
Meanwhile, given the limited number of entities in ESBM, we conduct $5$-fold 
cross-validation to reduce the risk of overfitting without losing the number of learning instances~\cite{Guo2016ADR}. 
Specifically, the entities in ESBM are divided into $5$ folds randomly. The parameters 
for each model are tuned on $4$-of-$5$ folds. The final fold in each case is utilized
to evaluate the optimal parameters. Since ESA has significantly better than all other state-of-the-art
methods in our baselines, we then compare the statistical significance among ESA and AutoSUMs (i.e.,
the original AutoSUM and the modified AutoSUM$^{1\sim5}$, respectively)
utilizing Student's paired t-test (\emph{p-value} $\leq 0.05$)~\cite{Hripcsak2005TechnicalBA}.

\subsubsection{Experimental Details}
For experimental details, 
we tune the parameters on a validation set (i.e., a part of the training
set). Specifically, to learn graph embeddings, we utilize TransE to pretrain the whole 
ESBM dataset. Here, the dimension of each triple is set to $100$. 
As for word embeddings,
we initialize the lookup table randomly,
where the dimension of each word is set to $100$.
Then we apply a BiLSTM with a single layer in each LSTM cell for feature extraction,
where the number of the layers in multi-aspect mechanism 
is set to $6$. 
In addition, the graph embedding of each triple is fixed 
after pretraining, while all other parameters in AutoSUM are initialized randomly and 
tuned without weight sharing. We train the AutoSUM model for $200$ epochs, and report the
results of the best epoch under early stopping.
\begin{table}
\scriptsize
\caption{F-measure comparison for top-$5$ and top-$10$ entity summarization.
$\uparrow\%$ is the relative improvement of AutoSUM,
and (+/-) is the indicator of significant improvement or degradation with respect to ESA
(\emph{p-value} $\leq 0.05$).}
\centering
\begin{tabular}{p{1.8cm}|p{1.0cm}|p{1.0cm}|
    p{1.0cm}|p{1.0cm}|p{1.0cm}|
    p{1.0cm}|p{0.7cm}|p{0.7cm}|p{0.7cm}}
\toprule[0.5pt]
\hline
\multirow{2}{*}{Model}
&
\multicolumn{2}{c|}{\textbf{DBpedia}} &
\multicolumn{2}{c|}{\textbf{LinkedMDB}} &
\multicolumn{2}{c|}{\textbf{ALL}}&
\multicolumn{3}{c}{$\uparrow\%$} \\
\cline{2-10}
& \textbf{k = 5} & \textbf{k = 10} & \textbf{k = 5} & \textbf{k = 10} & \textbf{k = 5} & \textbf{k = 10} 
& \textbf{min} & \textbf{max} & \textbf{avg} \\
\hline
RELIN & 0.242 & 0.455 & 0.203 & 0.258 & 0.231 & 0.399 & 25 & 118 & 72 \\
DIVERSUM & 0.249 & 0.507 & 0.207 & 0.358 & 0.237 & 0.464 & 12 & 114 & 54 \\
CD  & 0.287 & 0.517 & 0.211 & 0.328 & 0.252 & 0.455 & 10 & 110 & 52 \\
FACES & 0.270 & 0.428 & 0.169 & 0.263 & 0.241 & 0.381 & 23 & 162 & 73 \\
FACES-E & 0.280 & 0.488 & 0.313 & 0.393 & 0.289 & 0.461 & 17 & 48 & 38 \\
LINKSUM & 0.274 & 0.479 & 0.140 & 0.279 & 0.236 & 0.421 & 18 & 216 & 80 \\
MPSUM & 0.289 & 0.510 & 0.270 & 0.380 & 0.301 & 0.479 & 11 & 64 & 35 \\
\hline
ESA & 0.310	& 0.525	& 0.320 & 0.403 & 0.312 & 0.491 & 8 & 38 & 26 \\
\hline
AutoSUM & \textbf{0.387\bm{$^+$}} & \textbf{0.569\bm{$^+$}} & \textbf{0.443\bm{$^+$}} & \textbf{0.556\bm{$^+$}} 
& \textbf{0.403$^+$} & \textbf{0.565$^+$} & - & - & - \\
AutoSUM$^1$ & 0.303$^-$ & 0.425$^-$ & 0.316     & 0.442$^-$ & 0.290$^-$ & 0.462$^-$ & 22 & 40 & 31 \\
AutoSUM$^2$ & 0.316$^+$ & 0.538     & 0.375$^+$ & 0.463$^-$ & 0.333$^-$ & 0.517$^+$ & 6  & 22 & 16 \\
AutoSUM$^3$ & 0.221$^-$ & 0.390$^-$ & 0.330$^+$ & 0.406$^-$ & 0.252$^-$ & 0.394$^-$ & 34 & 75 & 49 \\
AutoSUM$^4$ & 0.254$^-$ & 0.417$^-$ & 0.309     & 0.394$^-$ & 0.270$^-$ & 0.411$^-$ & 36 & 52 & 43 \\
AutoSUM$^5$ & 0.325$^+$ & 0.532$^+$ & 0.343$^-$ & 0.413$^+$ & 0.323     & 0.502$^+$ & 7  & 35 & 21 \\
\hline
\bottomrule[0.5pt]
\end{tabular}
\end{table}

\subsection{Experimental Results}
As shown in Table 1 and 2, 
AutoSUM is significantly better than some existing state-of-art methods in our baselines.

\textbf{Comparison with Traditional Methods:} 
Compared with traditional methods
depending on manual feature extraction and multi-user preference simulation,
AutoSUM automates the above processes without any human expertise effectively.
The average improvement of AutoSUM over the best outperforming traditional methods is $38\%$ and $36\%$,
in terms of F-measure and MAP, respectively.
\begin{table}
\scriptsize
\caption{MAP comparison for top-$5$ and top-$10$ entity summarization.
$\uparrow\%$ is the relative improvement of AutoSUM,
and (+/-) is the indicator of significant improvement or degradation with respect to ESA
(\emph{p-value} $\leq 0.05$).}
\centering
\begin{tabular}{p{1.8cm}|p{1.0cm}|p{1.0cm}|
    p{1.0cm}|p{1.0cm}|p{1.0cm}|
    p{1.0cm}|p{0.7cm}|p{0.7cm}|p{0.7cm}}
\toprule[0.5pt]
\hline
\multirow{2}{*}{Model}
&
\multicolumn{2}{c|}{\textbf{DBpedia}} &
\multicolumn{2}{c|}{\textbf{LinkedMDB}} &
\multicolumn{2}{c|}{\textbf{ALL}}&
\multicolumn{3}{c}{$\uparrow\%$} \\
\cline{2-10}
& \textbf{k = 5} & \textbf{k = 10} & \textbf{k = 5} & \textbf{k = 10} & \textbf{k = 5} & \textbf{k = 10} 
& \textbf{min} & \textbf{max} & \textbf{avg} \\
\hline
RELIN & 0.342 & 0.519 & 0.241 & 0.335 & 0.313 & 0.466 & 25 & 115 & 55 \\
DIVERSUM & 0.310 & 0.499 & 0.266 & 0.390 & 0.298 & 0.468 & 30 & 94 & 53 \\
CD  & - & - & - & - & - & - & - & - & - \\
FACES & 0.255 & 0.382 & 0.155 & 0.273 & 0.227 & 0.351 & 69 & 234 & 114 \\
FACES-E & 0.388 & 0.564 & 0.341 & 0.435 & 0.375 & 0.527 & 15 & 64 & 36 \\
LinkSUM & 0.242 & 0.271 & 0.141 & 0.279 & 0.213 & 0.345 & 68 & 267 & 132 \\
MPSUM & 0.386 & 0.568 & 0.351 & 0.435 & 0.349 & 0.532 & 14 & 47 & 30 \\
\hline
ESA & 0.392 & 0.582 & 0.367 & 0.465 & 0.386 & 0.549 & 11 & 41 & 23 \\
\hline
AutoSUM & \textbf{0.459\bm{$^+$}} & \textbf{0.647\bm{$^+$}} & \textbf{0.517\bm{$^+$}} & \textbf{0.600\bm{$^+$}} 
& \textbf{0.476\bm{$^+$}} & \textbf{0.633\bm{$^+$}} & - & - & - \\
AutoSUM$^1$ & 0.419$^-$ & 0.508$^-$ & 0.420$^+$ & 0.522$^+$ & 0.389$^-$ & 0.563     & 10  & 27 & 18 \\
AutoSUM$^2$ & 0.404     & 0.598$^-$ & 0.431$^+$ & 0.525$^+$ & 0.412$^-$ & 0.578$^+$ & 8   & 20 & 14 \\
AutoSUM$^3$ & 0.291$^-$ & 0.456$^-$ & 0.383$^+$ & 0.488$^+$ & 0.317$^-$ & 0.465$^-$ & 23  & 58 & 41 \\
AutoSUM$^4$ & 0.333$^-$ & 0.486$^-$ & 0.376$^-$ & 0.467     & 0.346$^-$ & 0.480$^-$ & 28  & 38 & 34 \\
AutoSUM$^5$ & 0.405$^+$ & 0.582     & 0.368     & 0.473     & 0.412$^+$ & 0.550     & 11  & 40 & 21 \\
\hline
\bottomrule[0.5pt]
\end{tabular}
\end{table}

\textbf{Comparison with Deep Learning Methods:}
Compared with ESA,
which calculates attention scores without feature extraction and multi-user preference,
AutoSUM achieves the state-of-the-art performance.
The average improvement of AutoSUM over ESA is $26\%$ and $23\%$, in terms of F-measure and MAP, respectively.

In addition, we track the attention scores of entity \emph{Triathlon}
(\emph{Triathlon\_at\_ the\_ 2000\_Summer\_Olympics\_Men's})
in user-phase attention, as shown in Figure 2. 
We can observe that the user-phase attention
simulates $3$ groups of user preference of the entity,
and
the entity-phase attention allocates high attention 
scores to users who prefer medal as well as event than property, 
which is in accordance with 
the preference of most users in real world.

\begin{figure}
\centering
\includegraphics[scale=0.35]{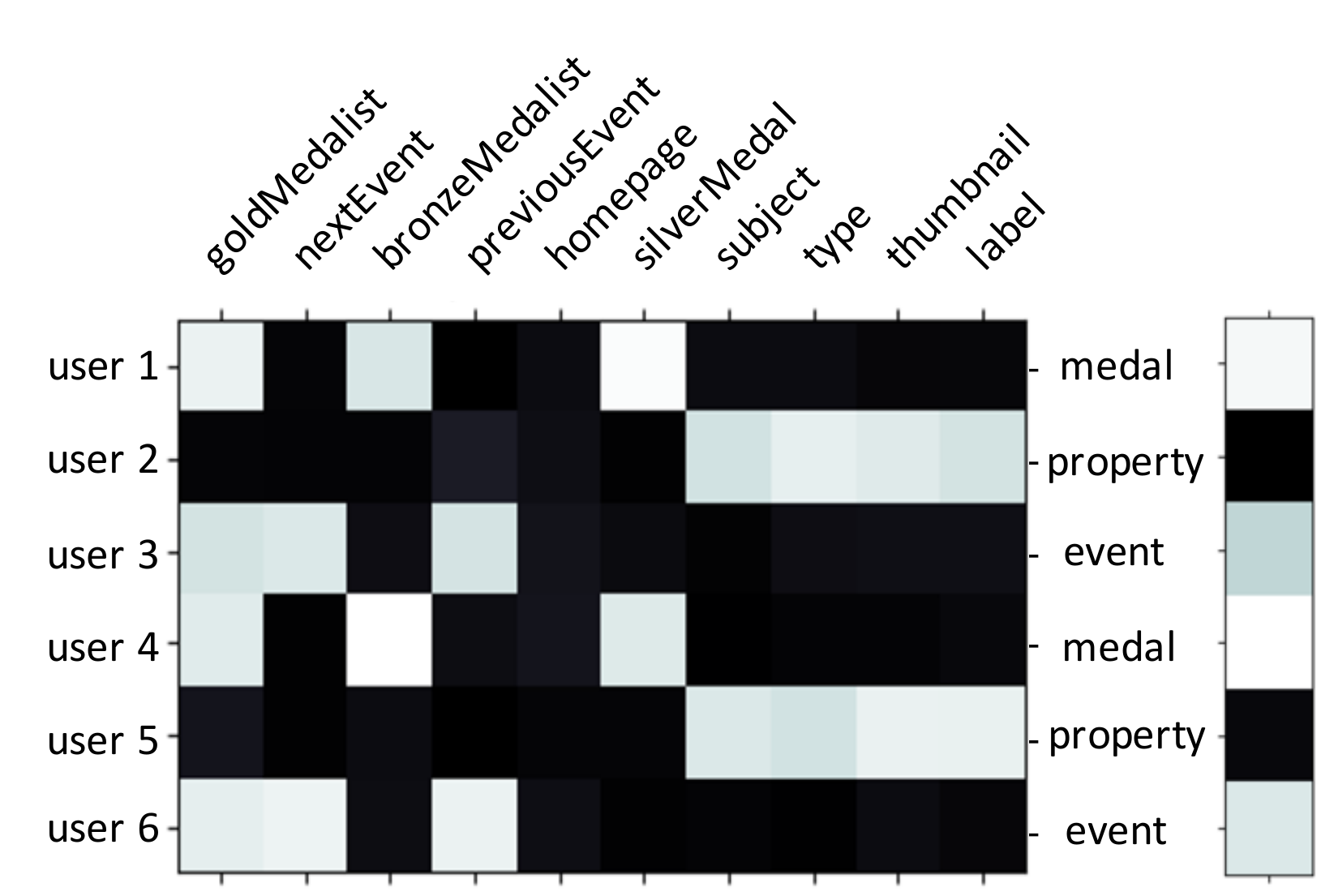}
\caption{The attention scores of
\emph{Triathlon\_at\_the\_2000\_Summer\_Olympics\_Men's.}}
\end{figure}

\subsection{Ablation Studies}
In this section, we provide ablation studies to demonstrate the effectiveness of the
primary modules in AutoSUM.

\textbf{AutoSUM$^{1}$:} 
To evaluate the features extracted by AutoSUM, 
AutoSUM$^{1}$ removes the BiLSTM in extractor and feeds the input representation 
of triples into simulator directly. Experimental results show the original AutoSUM is significantly 
better than AutoSUM$^{1}$, proving that the BiLSTM extracts high-quality features for
user-preference simulation.
      
\textbf{AutoSUM$^{2}$ and AutoSUM$^{4}$:} 
To explore whether the attention scores of different user preference are appropriate,
AutoSUM$^{2}$ removes the BiLSTM in simulator and allocates equal attention scores for
each user preference.
Meanwhile, we also attempt to replace the BiLSTM with an FCN, referred as Auto-SUM$^{4}$.
As shown in Table 1 and 2, the original AutoSUM gains a significant improvement
over AutoSUM$^2$ and AutoSUM$^4$, indicating the BiLSTM with a general attention function
allocates appropriate attention scores for each user preference. In addition,
we can observe that the 
performance of FCN (AutoSUM$^{2}$) is even worse than allocating equal attention scores 
(AutoSUM$^{4}$) in our experiments.

\textbf{AutoSUM$^3$:} 
For comparison, AutoSUM$^4$ removes the BiLSTM in both extractor and simulator. 
Experimental results show that the performance of Auto-SUM$^3$ is worse than AutoSUM$^1$ and
AutoSUM$^2$, which remove the BiLSTM in extractor and simulator respectively,
further proving the irreplaceable role of BiLSTM in AutoSUM.

\textbf{AutoSUM$^5$} To explore whether the multi-aspect mechanism 
captures the multi-aspect information from an entity,
we replace the multi-aspect mechanism with a single-aspect mechanism, i.e., setting 
the number of attention layers to $1$. 
As shown in Table 1 and 2, we can observe that the original AutoSUM outperforms AutoSUM$^5$ in
both F-measure and MAP. Experimental results indicate that the multi-aspect attention mechanism successfully 
captures the multi-aspect information. We also notice that AutoSUM$^5$ with a single-layer attention
mechanism still outperforms all other methods in our baselines including ESA. 

\section{Conclusion}
In this paper, we propose a novel integration model called AutoSUM to automate feature extraction 
and multi-user preference simulation for entity summarization. The performance of our proposed
AutoSUM is significantly better than other state-of-the-art methods in both F-measure and MAP. 
Meanwhile, sufficient ablation studies are provided to demonstrate the effectiveness of each module in AutoSUM. 
In the future, we expect to expand the ESBM dataset and introduce the notion of AutoSUM
into other applications such as recommender systems~\cite{DBLP:conf/pakdd/DoanYR19, pang2019novel}.

\section*{Acknowledgment}
This research is supported in part by the Beijing Municipal Science and Technology Project 
under Grant Z191100007119008.
%
%
%
%
%
%
\bibliographystyle{splncs04}
\bibliography{ref}
\end{document}